# Altermagnetic Metal-Organic Frameworks

*Diego López-Alcalá, Andrei Shumilin and José J. Baldoví\**

Instituto de Ciencia Molecular, Universitat de València, Catedrático José Beltrán 2, 46980 Paterna, Spain.

E-mail: j.jaime.baldovi@uv.es

■ **ABSTRACT**

Altermagnetism has recently emerged as a new class of spin compensated magnetic materials that exhibit momentum dependent spin splitting despite having zero net magnetization. The origin of these electronic signatures lies in symmetry operations that connect opposite spin sublattices while allowing spin splitting in momentum space. While most candidate materials identified so far belong to inorganic crystals with fixed lattice symmetries, the realization of altermagnetism ultimately requires platforms in which magnetic symmetry can be deliberately engineered. In this Perspective, we discuss how metal-organic frameworks (MOFs) provide a unique chemical platform to address this challenge. We first place altermagnetism in the broader context of magnetic and electronically active metal-organic networks, highlighting how reticular chemistry enables precise control over lattice geometry, dimensionality and electronic structure. We then discuss how these features position framework materials as promising candidates for realizing altermagnetism and highlight the key challenges that must be addressed to translate theoretical proposals into experimentally accessible systems. Finally, we critically assess current experimental challenges and outline emerging directions for realizing and controlling altermagnetism in coordination framework materials, which emerge as a versatile and powerful platform for exploring new paradigms in spintronics.

■ **INTRODUCTION**

**Why does altermagnetism need chemistry?**

Altermagnetism has been recognized as the scientific breakthrough in Physics 2024[1] and represents a fundamentally new form of magnetic order, redefining how spin polarization can arise in crystalline solids without net magnetization.[2–5] By demonstrating that symmetry alone can generate momentum dependent spin splitting, altermagnetism bridges key concepts from ferromagnetism, antiferromagnetism and spin-orbit physics, offering an attractive route toward spin transport and spintronic functionalities without macroscopic magnetic fields.[6] To date, however, the exploration of altermagnetism has been largely driven by symmetry analysis and first principles studies within a relatively narrow set of inorganic materials.[7]

This raises a central question: if altermagnetism is governed by symmetry, where can symmetry itself be deliberately designed rather than merely identified? Dense inorganic crystals offer limited freedom in this regard, as their lattice topology, electronic structure and magnetic symmetry are tightly constrained by atomic packing. In contrast, chemistry and coordination chemistry in particular, provides a fundamentally different paradigm. Through the modular combination of metal centers and organic linkers, chemical synthesis enables the deliberate construction of lattices, sublattices and magnetic motifs with programmable geometry, connectivity and electronic character.[8,9]

In this Perspective, we explore the potential of metal-organic frameworks (MOFs)– recognized with the Nobel Prize in Chemistry in 2025–[10,11] to provide ideal and largely untapped platforms for altermagnetism. Rather than searching for altermagnetism within a fixed materials landscape, coordination frameworks allow symmetry to be engineered by design, thus transforming altermagnetism from a rare emergent phenomenon into a chemically accessible magnetic state. We critically examine the current theoretical and experimental challenges that must be overcome to realize altermagnetism in MOFs and outline emerging directions that exploit their unique structural and electronic versatility. Therefore, this Perspective aims to clarify why altermagnetism needs chemistry and how chemical design may ultimately redefine the scope, robustness and functionality of altermagnetic materials.

## ■ ALTERMAGNETISM AS A SYMMETRY-DRIVEN PHENOMENON

Altermagnets form a distinct class of collinear magnetic materials with two magnetic sublattices, alongside ferrimagnetic (FiM) and antiferromagnetic (AFM) materials. These three classes are distinguished purely by symmetry considerations –specifically, by the presence and character of symmetries that relate the two sublattices–.[2] In the absence of any sublattice-relating symmetry, a material is classified as FiM, whose macroscopic properties closely resemble those of ferromagnets. A material is identified as AFM when such a symmetry exists and, upon acting on the electron momentum, either preserves it (e.g., a translation $t$) or reverses it (e.g., inversion $i$ or the combined $it$ operation). The third possibility defines the altermagnetic (AM) phase, where sublattice-relating symmetries exist but act on electron momentum in a more complex manner. Altermagnets combine hallmark properties of both antiferromagnets and ferromagnets.[7] Like AFM materials, they exhibit zero net magnetization, rendering them robust against external magnetic fields and feature linear magnon dispersion. At the same time, similar to ferromagnets, their spin-split electronic bands enable electrical detection of magnetic order parameters via the anomalous Hall effect[12,13] and tunneling magnetoresistance.[14] Moreover, spin injection into adjacent materials is possible due to spin-polarized currents arising from the spin splitting effect,[15] which is a strong analogue of the much weaker spin Hall effect driven by spin–orbit coupling (SOC).[16]

Usually, AM materials are further classified into $d$-, $g$- and $i$-wave symmetry classes, characterized by two, four and six planes, respectively, in the spin splitting of the electronic dispersion. While all altermagnets share zero net magnetization and non-relativistic band splitting, their functional responses differ markedly. Notably, only $d$-wave altermagnets support spin splitting currents in the linear (Ohmic) response regime, whereas in the other symmetry classes analogous effects emerge only as higher-order responses to an applied electric field.[17]

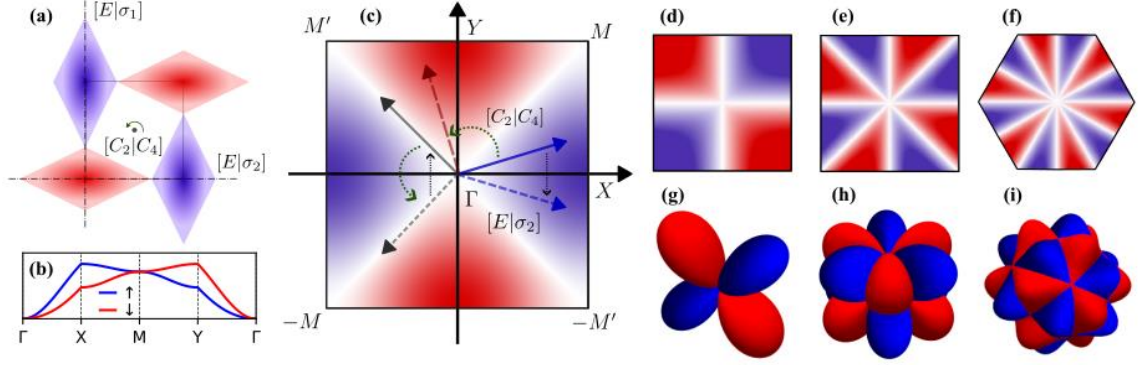

**Figure 1: (a)** Schematic illustration of a toy model of AM material belonging to the *d*-wave class. **(b)** Representative electron dispersion in such a material. **(c)** Diagram illustrating the action of symmetry operations on the electron dispersion: the $[C_2 \mid C_4]$ operation preserves the band structure with a spin-flip, indicated by the red dashed arrow; the $[E \mid \sigma_2]$ operation preserves the band structure without a spin-flip, indicated by the blue dashed arrow. The consecutive application of these two symmetry operations to a wavevector along the $\Gamma M'$ direction preserves momentum while including a spin-flip, thereby preventing electron band splitting along this direction. **(d-f)** Representative band-splitting patterns in 2D *d*-, *g*- and *i*-wave altermagnets. **(g–i)** Representative angular dependences of the band splitting in 3D *d*-, *g*- and *i*-wave altermagnets.

Because altermagnets are defined by non-relativistic electronic properties, it is conventional to describe AM materials using non-relativistic symmetry notation. In this description, each relevant symmetry operation is written as a pair $[E \mid g_0]$ or $[C_2 \mid g_{sf}]$. Here, the first symbol denotes the operation in spin space, where $E$ represents spin conservation and $C_2$ a 180° spin rotation (spin flip), while the second symbol specifies the corresponding operation acting on the crystal lattice. The coexistence of these two types of symmetry operations —$[E \mid g_0]$ and $[C_2 \mid g_{sf}]$— is what defines an AM state itself and determines its specific symmetry class. To illustrate this idea, Figure 1a shows a simple 2D model of an altermagnet composed of two magnetic sublattices.[18] Each sublattice is formed by a set of orbitals related by a 90° rotation and occupied by electrons with opposite spin orientations. In this example, the $[C_2 \mid g_{sf}]$ operation corresponds to a 90° rotation about the *z*-axis ($C_4$), which maps one sublattice onto the other while reversing the spin. By contrast, two mirror planes, $\sigma_1$ and $\sigma_2$, act as $[E \mid g_0]$ symmetries, preserving the spin orientation.

Figure 1b shows the resulting electronic band structure. Owing to the AM symmetry, the bands exhibit a wavevector-dependent spin splitting that vanishes along specific high-symmetry directions. In 3D systems, these directions generalize into nodal planes. Such spin splitting patterns are often summarized using schematic diagrams, as shown in Figure 1c.[19] In these diagrams, regions of the First Brillouin zone with opposite sign of spin splitting are indicated by different colors. Applying a $[C_2 \mid g_{sf}]$ operation to a given wavevector maps the electronic state onto one with the same energy but opposite spin, corresponding to a transition between regions of opposite color (red dashed arrow in Figure 1c). In contrast, a $[E \mid g_0]$ operation preserves both the energy and the spin, leading to a transition within regions of the same color (blue dashed arrow). Finally opposite wavevectors always belong to the same color region in the schematic representation. Along certain high-symmetry directions, specific symmetry operations that include a spin flip can map a wavevector onto itself, giving rise to nodal directions and planes. This is illustrated in Figure 1c, where the black arrow denotes a wavevector along the $\Gamma M'$ direction. In this case, the wavevector remains invariant under the successive application of the $[\,C_2\,|\,C_4\,]$ and $[\,E\,|\,\sigma_2\,]$ symmetry operations.

The wide variety of possible material symmetries gives rise to distinct angular dependences of spin splitting in altermagnets. In Figure 1d–f, we present schematic diagrams of the most common spin splitting patterns in the first Brillouin zone of 2D $d$-, $g$- and $i$-wave altermagnets. Figure 1g–i shows the typical angular dependences of the electron-band spin splitting in 3D $d$-, $g$- and $i$-wave altermagnets.

## ■ MAGNETIC MOFs AS SYMMETRY-ENGINEERED QUANTUM MATERIALS

### General background of magnetic MOFs for spintronics

Over the past decade, coordination frameworks have evolved from chemically versatile porous materials into genuinely programmable platforms for controlling electronic and magnetic degrees of freedom.[20,21] Their defining strength lies in the intimate coupling between ligand symmetry, coordination geometry and orbital character, which enables a level of structural and electronic diversity that is difficult to achieve in conventional crystalline solids.[22–24] Within this context, magnetic MOFs and coordination polymers have emerged as particularly powerful systems for engineering spin dependent functionalities through chemical design.[25–28] A growing body of experimental work has demonstrated that organic ligands can efficiently mediate magnetic exchange between metal centers, stabilizing cooperative magnetic behavior even in low density and low dimensional lattices.[29]

The first generation of magnetic and electrically conductive MOFs established the conceptual and chemical foundations for coordination frameworks as multifunctional magnetic materials. A seminal breakthrough was achieved with the incorporation of redox-active quinoid ligands into extended frameworks, exemplified by the layered iron–chloranilate system $(Me_2NH_2)_2[Fe_2L_3]\cdot 2H_2O\cdot 6DMF$ (L = 2,5-dichloro-3,6-dihydroxy-1,4-benzoquinone), which demonstrated spontaneous magnetization with a Curie temperature ($T_C$) of 80 K in its solvated form and $T_C$ = 26 K upon desolvation.[30] This work provided the first structurally characterized example of a microporous magnet incorporating tetraoxolene radical ligands and highlighted the decisive role of metal–radical exchange in overcoming the intrinsically weak magnetic coupling typical of MOFs (Figure 2a). Building on this concept, heterogeneous redox chemistry was subsequently exploited to further amplify magnetic interactions in two-dimensional (2D) iron–quinoid frameworks. In particular, single-crystal-to-single-crystal reduction of the same chloranilate-based framework yielded the radical-rich phase $[Fe^{III}_2(L^{3-\bullet})_3]^{3-}$, exhibiting long-range magnetic order up to $T_C$ = 105 K together with measurable electrical conductivity ($\sigma \approx 10^{-3} - 10^{-2}$ S cm$^{-1}$), representing one of the highest ordering temperatures reported for a MOF to date.[31] In parallel, an alternative strategy based on extended π–d conjugation emerged with the development of layered two-dimensional MOFs such as the perthiolated coronene–iron framework (PTC–Fe), which combines semiconducting transport with room-temperature conductivity approaching ~10 S cm$^{-1}$ and ferromagnetic (FM) correlations below ~20 K, mediated by delocalized π electrons coupling intermediate-spin Fe(III) centers.[32] In the 2D manganese benzoquinoid framework (Me$_4$N)$_2$[Mn$^{2+}$$_2$(L$^{2-}$)$_3$], post-synthetic reduction to generate semiquinoid radical ligands transforms a paramagnetic lattice into a long-range ordered magnet with a Curie temperature of $T_C$ = 41 K, accompanied by a striking 200 000-fold increase in electrical conductivity.[33] Together, these pioneering studies established redox-active ligands and π–d conjugation as two central chemical design principles for simultaneously enabling magnetism and charge transport in MOFs, setting the stage for subsequent generations of framework-based spintronic and AM materials.

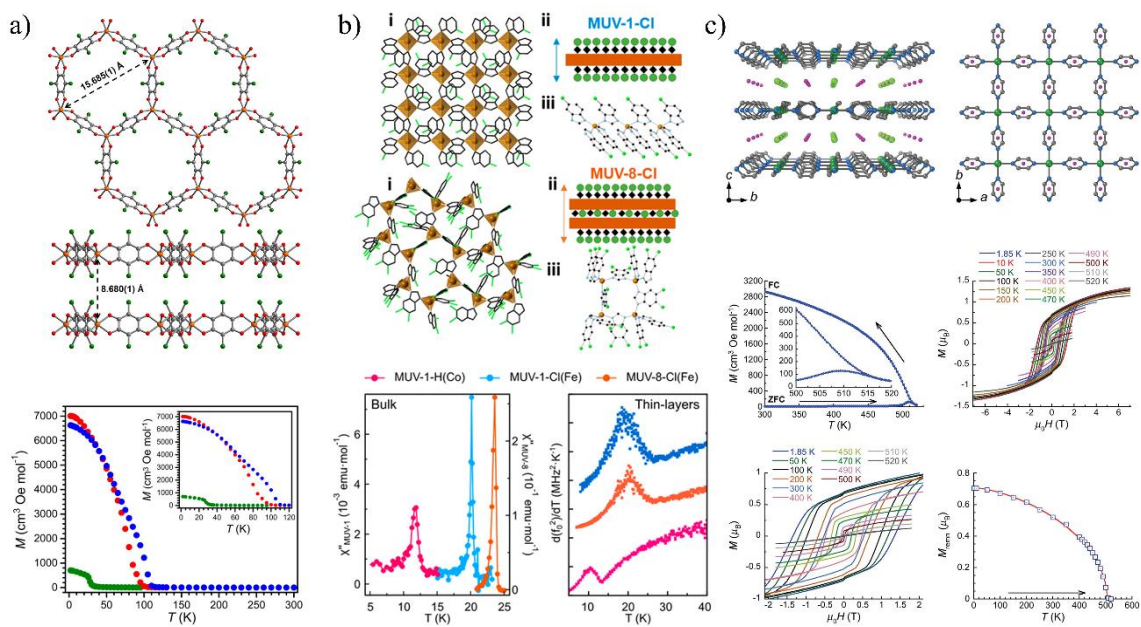

**Figure 2: (a)** X-ray crystal structure of [Fe$_2$L$_3$]$^{3-}$, as observed in 2, viewed along the crystallographic *c* axis and *b* axis with selected Fe···Fe distances. Orange, green, red, and gray spheres represent Fe, Cl, O, and C atoms, respectively; cations and DMF molecules are omitted for clarity (top). Variable-temperature field-cooled magnetization data for 1 (red), 1a (green), and 2 (blue). Reproduced with permission from REF. 30, American Chemical Society. **(b)** Comparison of the crystal structures of MUV-1-Cl (a) and MUV-8-Cl (b). Structure of a single layer (i) viewed along the *c* axis (*ab* plane). Schematic representation of a single layer (ii), showing an increase thickness for the "double layer" of irons in the MUV-8-Cl (from 1 to 1.5 nm). Coordination modes of the ligands (iii) (top). Comparison of the out-of-phase a.c. susceptibility signals for bulk systems (left) and the derivative of $f_0^2(T)$ as a function of temperature for thin layers (right) (bottom). Reproduced with permission from REF. 34, American Chemical Society.

**(c)** Perspective view (along the a direction) of 2·0.25(THF) showing the alternation of $Li_{0.7}Cl_{0.7}$ and neutral $Cr^{II}(pyz^{•-})_2$ layers stacking along the c direction. Eclipsed layered structure viewed along the c direction. Cr is shown in dark green, N in blue, and C in dark gray. Cl (light green) and Li (pink) are shown at a fixed occupancy of 70% according to the elemental composition. Hydrogen atoms have been omitted for clarity. Zero field-cooled (ZFC) and field-cooled (FC) magnetization data obtained under an applied dc magnetic field of 50 Oe at 5 K min$^{-1}$. Inset shows a magnified view of the main plot in the 500 to 520 K temperature range. The solid lines are a guide for the eye. Magnetization versus applied dc magnetic field data (at 5 to 12 Oe s$^{-1}$) in the −7 to 7 T field range, from 1.85 to 520 K (42). Magnified view of selected data in the −2.1 to 2.1 T field range. Temperature dependence of the remnant magnetization, determined from the M versus m0H data between 1.85 and 520 K. The solid red line is the best fit to the mean-field (MF). Reproduced with permission from REF. 43, Science AAAS.

A second generation of magnetic MOFs has been realised in recent years, marked by the stabilization of long-range magnetic order in low-dimensional and atomically defined architectures together with new electronic and spin functionalities. A paradigmatic example is provided by the family of layered MOFs MUV-1(M) (M = Fe, Co, Mn, Zn), constructed from benzimidazole-type ligands and divalent metal centers, which crystallize as van der Waals (vdW)-bonded magnetic layers that can be mechanically exfoliated down to few-layer thickness. These materials retain cooperative magnetic order upon exfoliation, with Néel temperature ($T_N$) of 20 K for MUV-1(Fe) and MUV-1(Mn), demonstrating that intrinsic magnetism can persist in ultrathin molecular crystals (Figure 2b).[34,35] Moving to the strict 2D limit, on-surface synthesis has enabled atomically thin metal-organic coordination networks with fully cooperative magnetic behavior.[36] In a landmark study, Fe-9,10-dicyanoanthracene (Fe-DCA) networks grown on Au(111) were shown to exhibit extended FM order with an out-of-plane easy axis, large coercive fields exceeding 2 T and a $T_C \approx 35$ K, despite only ~5% of the lattice sites being occupied by magnetic Fe atoms.[37] Complementing these systems, mixed-valence and mixed-spin two-dimensional frameworks such as $Fe_2(Fe-DPyP)_3$, assembled from 5,15-di(4-pyridyl)-10,20-diphenylporphyrin ligands, realize chemically distinct iron sites with spin states $S = 1$ and $S = 3/2$ arranged on intertwined Kagome and honeycomb sublattices, leading to an in-plane FM ground state directly resolved by scanning tunneling spectroscopy.[38] Finally, conjugated 2D MOFs based on square-planar metal nodes and redox-active ligands have enabled electrically conductive magnetic frameworks with room-temperature conductivities spanning several orders of magnitude and magnetically ordered phases stabilized through extended π-d conjugation.[39] Collectively, these results define a new generation of magnetic MOFs in which long-range magnetic order, reduced dimensionality, electronic transport and chemically encoded lattice complexity coexist within a single materials platform, setting the stage for symmetry-driven magnetic phenomena such as altermagnetism.[40]

Among the wide variety of magnetic MOFs reported to date, systems based on redox-active ligands have consistently emerged as the most successful platform for achieving high magnetic ordering temperatures and spintronic-relevant functionalities.[27] A paradigmatic example is provided by coordination networks incorporating organic radical linkers, most notably tetracyanoethylene ([TCNE]•−), which enabled the realization of molecule-based magnets such as V[TCNE]$_x$ (x ≈ 2) with $T_C$ approaching 400 K, establishing the long-standing record for organic and metal–organic magnetic materials.[41] Building on this principle, a major advance was achieved with layered chromium–pyrazine (pyz) frameworks, CrCl$_2$(pyz)$_2$, where mixed-valence pyz radicals mediate strong exchange between Cr centers, resulting in ferrimagnetic order below $T_C$ ≈ 55 K together with measurable electrical conductivity.[42] Strikingly, post-synthetic reduction of structurally related Cr–pyz coordination networks has recently pushed this strategy to its limit, yielding lightweight metal–organic ferrimagnets with critical temperatures of 242 K in the pristine phase and exceeding 500 K following post-synthetic modification, accompanied by large room-temperature coercivities of up to 7500 Oe (Figure 2c).[43,44] These results unambiguously demonstrate that metal-radical exchange, enabled by redox-active ligand scaffolds, is the only strategy that has so far allowed MOFs to reach magnetic energy scales comparable to those of inorganic magnets, positioning redox-active frameworks as the most promising candidates for future high-temperature magnetic and spintronic applications.

Beyond simple magnetic order, coordination frameworks represent a uniquely versatile platform for realizing complex magnetic lattices in which multiple chemically distinct spin sublattices coexist within a single material. 2D metal-organic networks, realized either as layered bulk phases or surface-supported architectures, naturally host inequivalent magnetic sites with different spin states, mixed-valence configurations and intertwined lattice geometries. These features give rise to rich spin excitation spectra and competing FM, FiM and AFM interactions that extend well beyond conventional nearest-neighbor superexchange. Importantly, this intrinsic magnetic complexity is encoded chemically through the choice of metal centers, ligands and connectivity, enabling the deliberate construction of magnetic substructures that are exceptionally difficult to engineer in dense inorganic crystals. Despite these advances, magnetic coordination frameworks have so far been predominantly explored within conventional magnetic paradigms, leaving a broad space of unconventional magnetic states largely unexplored.

**Framework chemistry as a symmetry engine**

A defining feature of MOFs is their exceptional versatility in realizing and manipulating symmetry. In coordination frameworks, symmetry is not imposed by dense atomic packing, but deliberately constructed through the choice of metal nodes, organic linkers and their modes of connectivity. As a result, MOFs and related coordination networks provide access to an unusually rich symmetry landscape, encompassing a broad range of point groups, lattice symmetries and dimensionalities within a chemically unified materials platform.[45–48]

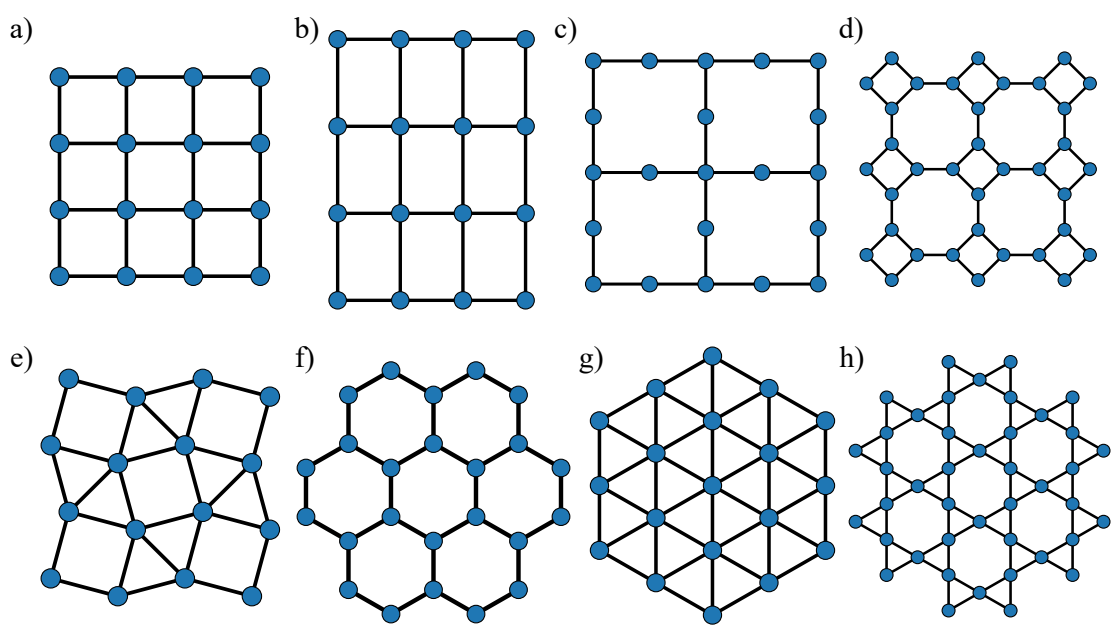

**Figure 3:** Representative 2D topologies hosting magnetic sublattices in 2D MOFs. The diagrams illustrate the nodes (blue spheres) and linkers (black lines) corresponding to the RCSR codes and Schläfli symbols: **(a)** sql ($4^4$), **(b)** rectangular distortion of the sql net, **(c)** Lieb lattice, **(d)** fes ($4.8^2$), **(e)** tts (snub square, 3.3.4.3.4), **(f)** hcb (honeycomb, $6^3$), **(g)** hxl ($3^6$) and **(h)** kgm (Kagome, 3.6.3.6). These geometries exhibit distinct symmetry-breaking features and sublattice interplays crucial for emerging AM phases.

This versatility originates from the modular nature of coordination chemistry. Organic linkers with well-defined geometry, rigidity and connectivity act as symmetry defining elements that propagate their local point symmetry into the extended lattice.[49] By varying linker shape, connectivity and binding directionality, coordination frameworks can be rationally assembled into square, rectangular, honeycomb, Kagome or hexagonal, as well as more complex topologies (Figure 3).[50,51] These chemically accessible lattice symmetries have enabled coordination frameworks to host a broad range of modern magnetic and electronic phenomena, each closely linked to specific network topologies. Square and Lieb lattices have been shown to support robust long-range magnetic order in layered MOFs and coordination polymers.[52,53] Honeycomb lattices have enabled Dirac like electronic dispersions in conjugated frameworks, opening routes toward topological and relativistic electronic behavior.[54] Kagome networks, in contrast, naturally promote geometric magnetic frustration, providing model platforms for spin liquid states and flat band physics.[55,56] Cairo or pentagonal lattices introduce further frustration and quasiperiodic magnetic motifs, giving rise to exotic magnetic ground states in inorganic materials, including complex noncollinear order and quasicrystalline approximants (Figure 4a). Importantly, this topology has recently been realized in MOFs through isoreticular design based on Cairo pentagonal tiling, demonstrating that such unconventional lattices are also synthetically accessible within coordination chemistry.[57] Finally, mixed or low-symmetry lattices provide fertile ground for unconventional magnetic order in coordination frameworks. Archimedean and distorted tessellations realized in MOFs have been shown to host multiple inequivalent magnetic sites and competing exchange pathways, leading to noncollinear spin configurations and frustration-driven magnetic behavior (Figure 4b).[58] The modularity of coordination frameworks has also recently enabled the incorporation of heavier 4d transition metals, further expanding the chemical space accessible to magnetic MOFs.[59] These results highlight how chemically encoded low-symmetry architectures enable magnetic textures that are rarely accessible in conventional inorganic crystals.[60] Many of these lattice topologies, remain rarely accessible in classical inorganic materials, highlighting the unique role of coordination chemistry as a symmetry engine for unconventional magnetic states.

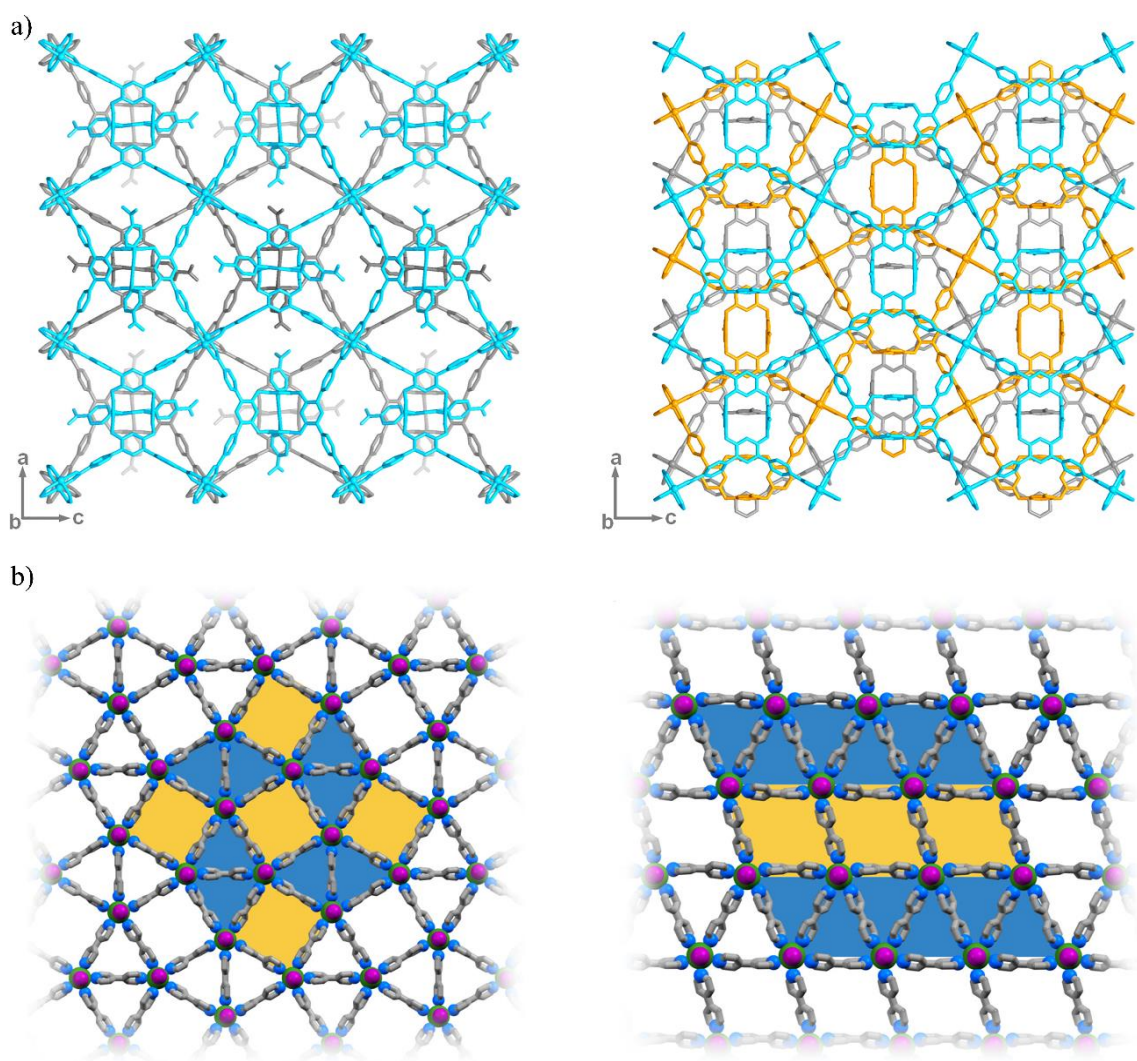

**Figure 4: (a)** Views of stacking layers in mcm-MOF-1(left) and mcm-MOF-2 (right). Reproduced with permission from REF. 57, CellPress. **(b)** Single-crystal X-ray structure of Gd (left). Single-crystal structure of the trace impurity phase Gd′ (right). Color codes: Gd, green; I, purple; N, blue; C, gray. H atoms and cocrystallized $CH_3CN$ molecules have been omitted. Reproduced with permission from REF. 58, American Chemical Society.

Finally, the inherent softness and modularity of coordination frameworks allow symmetry to be modified well beyond the initial synthesis step. Experimental studies on layered magnetic coordination polymers and 2D MOFs have shown that pre-synthetic ligand functionalization, guest inclusion, redox chemistry and controlled post-synthetic transformations can systematically alter local coordination environments, interlayer registry and effective symmetry, while preserving crystallinity, magnetic order and the underlying network topology.[61] This ability to generate isoreticular families of closely related frameworks, in which symmetry, electronic structure and magnetic interactions can be tuned in a controlled and chemically encoded manner, is essentially inaccessible in rigid inorganic crystals.[35] In the context of altermagnetism, such chemical agility represents a unique opportunity to iteratively explore and refine symmetry conditions, transforming symmetry from a fixed structural constraint into a continuously adjustable design variable.

Taken together, these features establish MOFs as a chemically programmable platform in which symmetry can be explored, tuned and deliberately engineered across an exceptionally broad parameter space. This intrinsic symmetry versatility provides the conceptual foundation for realizing unconventional magnetic states in framework materials and establishes altermagnetism as a chemically programmable property rather than a constraint imposed by existing crystal lattices.

## ■ EMERGENCE OF ALTERMAGNETISM IN MAGNETIC MOFs

### Altermagnetism meets framework materials

Initial theoretical efforts have appeared only recently, demonstrating that the symmetry requirements for altermagnetism can be fulfilled within coordination frameworks through deliberate lattice design. In a seminal contribution, Che *et al.* proposed the first realization of AM spin splitting in MOFs by engineering tetragonal lattices based on $M(pyz)_2$ networks, with M = Ca and Sr (Figure 5a).[62] In these systems, an effective reduction of the ligand scaffold induces unconventional momentum-dependent spin splitting in the absence of net magnetization. This behavior originates from the symmetry-enforced coupling between magnetically compensated sublattices through the combined spatial and spin operation $\left[ C_2 \mid C_{4_z} \right]$, which gives rise to a characteristic *d*-wave AM anisotropy.

**Table 1:** Structure type, AM anisotropy, spin splitting ($E_S$, in meV) and Néel temperature ($T_N$, in K) calculated in AM MOFs reported in bibliography.

|  | Structure type | AM anisotropy | $E_s$ | $T_N$ | Ref. |
|---|---|---|---|---|---|
| Ca(pyz)$_2$ | Monolayer | d-wave | - | 15.5 | 62 |
| Sr(pyz)$_2$ | Monolayer | d-wave | - | 11.5 | 62 |
| Cr(DAind)$_2$ | Monolayer | d-wave | 84 | 25 | 63 |
| Cr(diz)$_2$ | Monolayer | d-wave | 16 | 112 | 65 |
| Cr(c-pyr)$_2$ | Monolayer | d-wave | 30 | 177 | 65 |
| Cr(f-pid)$_2$ | Monolayer | d-wave | 37 | 183 | 65 |
| t-Cr$_2$(Pyc-O)$_8$ | Monolayer | d-wave | 184 | 57 | 66 |
| Cr(imz)$_2$ | Monolayer | g-wave | 65 | 37 | 63 |
| Cr(Dapent)$_2$ | Monolayer | g-wave | 30 | 58 | 63 |
| Ru2(TCNQ)$_2$ | Monolayer | g-wave | 100 | 24 | 64 |
| Cr(tcb)$_2$ | Bilayer | d-wave | 162 | 205 | 69 |
| Cr(hcb)$_2$ | Bilayer | d-wave | 314 | 85 | 69 |
| [C(NH2)$_3$]Cr(HCOO)$_3$ | Bulk | d-wave | 20 | - | 67 |

Building on this lattice-specific realization, López-Alcalá et al. subsequently proposed a more general design methodology for altermagnetism in 2D MOFs based on ligand symmetry.[63] In this approach, the use of non-centrosymmetric organic linkers is shown to systematically lower the symmetry of the metal-organic network in a manner that fulfills the symmetry requirements for altermagnetism. Starting from pyz-based lattices, the replacement by imidazole-based (imz) ligands induces an explicit symmetry breaking between magnetic sublattices, resulting in a g-wave AM spin splitting (Figure 5c). At the microscopic level, ligand symmetry enables the emergence of specific symmetry operations that connect spin-compensated sublattices. In particular, the combined operations $[E | C_{4z}]$ and $[C_2 | g]$, act as the symmetry elements responsible for enforcing AM spin splitting in momentum space. This analysis explicitly demonstrates how chemical control over ligand symmetry directly translates into the magnetic spin space group of the framework. Crucially, this concept is not restricted to a specific ligand but can be extended to a broader class of non-centrosymmetric organic linkers, thereby opening a general and chemically intuitive route to tune the electronic and magnetic properties of AM MOFs. Moreover, the authors demonstrate that frontier molecular orbital engineering (FMOE) provides an additional degree of control over AM anisotropy. By selectively modifying the symmetry and energetic alignment of ligand frontier orbitals, a transition from g-wave to d-wave AM anisotropy can be achieved, since $[C_2 | g C_4]$ connects spin sublattices. This highlights orbital design as a powerful tool for accessing distinct AM phases within a single framework platform. Following this symmetry-driven design methodology, Zhang et al. proposed Cairo-type pentagonal 2D MOFs as a platform for AM spin splitting.[64] First-principles calculations identified Ru$_2$(TCNQ)$_2$ as a g-wave altermagnet exhibiting sizable nonrelativistic spin splitting while remaining fully magnetically compensated. In this system, altermagnetism is protected by combined symmetry operations of the form $[C_2 | M]$, where mirror operations connect antiparallel spin sublattices within the pentagonal lattice, enforcing momentum-dependent spin splitting without net magnetization.

In complementary work, FMOE has been employed to directly stabilize an AM ground state in tetragonal Cr-based MOFs, where ligand-centered spin polarization provides the symmetry breaking required for altermagnetism.[65] In this approach, the AM phase is driven by the symmetry and energetic alignment of ligand frontier orbitals hybridized with metal $d$ states, rather than by lattice geometry alone. Crucially, overall magnetic compensation is preserved whereas ligand spin polarization connects magnetic sublattices via $[C_2 \mid S_{4_z}]$. As a result, sizable momentum-dependent spin splitting is obtained together with predicted magnetic ordering temperatures approaching 180 K, highlighting the potential of orbital design to achieve robust AM phases in coordination frameworks.

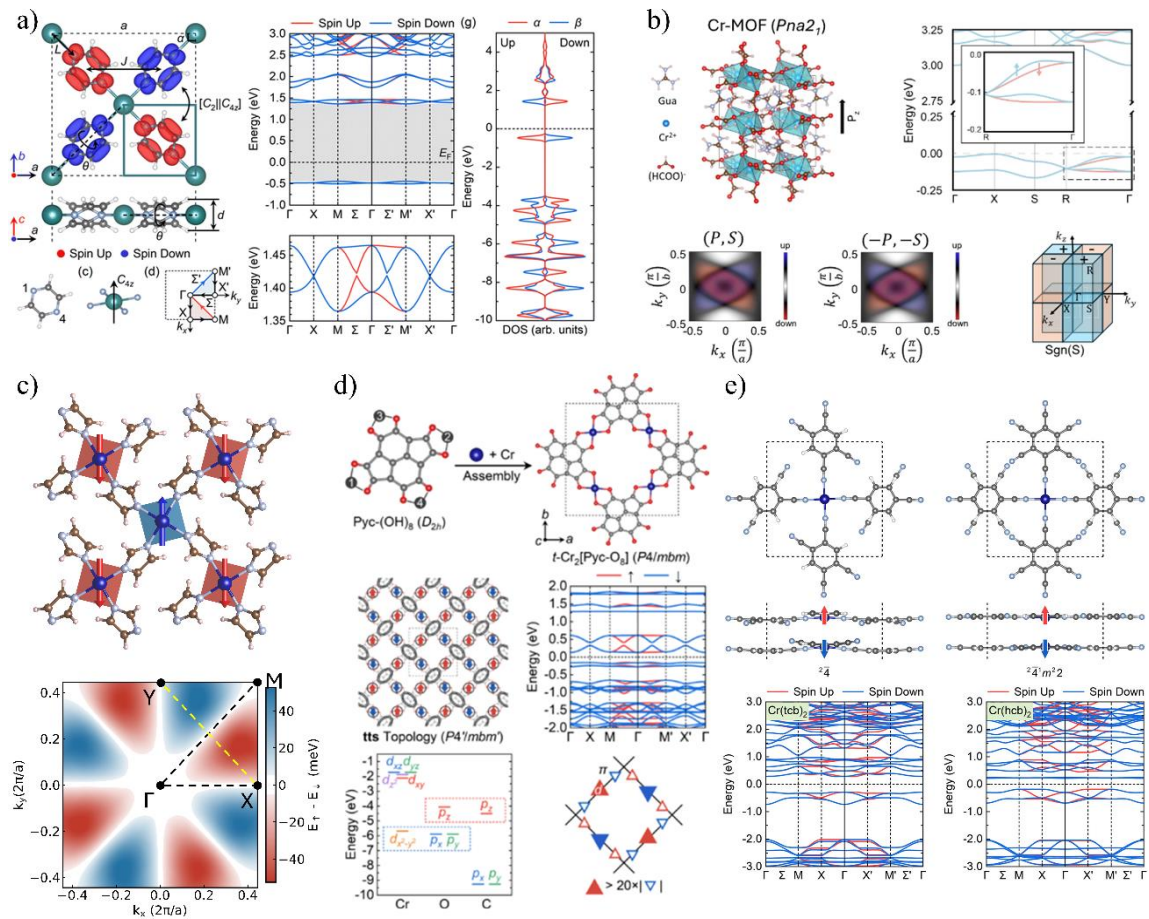

**Figure 5: (a)** Altermagnetic 2D Ca(pyz)$_2$ with unconventional spin-splitting and valley-dependent spin-polarized sublattices. Structure and spin valleys of the altermagnetic 2D Ca(pyz)$_2$. Green, gray, blue, and white balls represent Ca, C, N, and H atoms, respectively. The first Brillouin zone (BZ) of the altermagnetic 2D Ca(pyz)$_2$ with high symmetry $k$-points and $k$-paths. The red and blue arrows represent the different spin-splitting paths and the corresponding triangle areas are the two irreducible parts of the first BZ. Band structure of the altermagnetic 2D Ca(pyz)$_2$ with HSE06 functional. Sublattice and spin resolved density of states (DOS) of the altermagnetic 2D Ca(pyz)$_2$. Reproduced with permission from REF. 62, American Chemical Society. **(b)** Crystal structure of Cr-MOF. Spin-split band structure of Cr-MOF. Spin-resolved isoenergy contour ( $E=-0.05$ eV) for the (*P*,*S*) state at the $k_z=0.25$ plane, blue and red denote the spectral weight for spin-up and spin-down, respectively. Same as previous image, but for the (−*P*,−*S*) state. The sign of $\Delta E_k^S$ for the valence band maximum in reciprocal space. Reproduced with permission from REF. 67, American Physical Society. **(c)** Representative imz-based 2D MOF antiferromagnetically coupled. Spin splitting in VBM along 2D MOF plane. Reproduced with permission from REF. 63. **(d)** Assembly of *t*-Cr$_2$[Pyc-O$_8$] from Cr and Pyc-(OH)$_8$ molecules. Blue, gray, and red balls represent Cr, C, and O atoms, respectively. H atoms are omitted for clarity. The tts net in the altermagnetic *t*-Cr$_2$[Pyc-O$_8$]. Band structure of monolayer *t*-Cr$_2$[Pyc-O$_8$]. Orbital energy level matching of Cr, O around Cr, and C directly connected with O. Magnetic exchange pathway of *d*-π-π-*d* between two adjacent Cr atoms. Reproduced with permission from REF. 66, American Physical Society. **(e)** Crystal and band structures of bilayer MOF composed of chromium and cyanobenzene. Dark blue, gray, light blue, and white balls represent Cr, C, N, and H atoms, respectively. Reproduced with permission from REF. 69, American Chemical Society.

More recently, AM states in MOFs have been proposed from a fully symmetry-based and mathematical perspective. In this approach, all square tessellations compatible with 2D coordination networks were systematically screened to identify lattice symmetries capable of fulfilling the symmetry breaking requirements for altermagnetism[66]. As a proof of concept, the framework Cr$_2$[Pyc-O$_8$] was proposed as a 2D d-wave altermagnet, in which antiparallel spins occupying lower-symmetry Wyckoff positions generate time-reversal symmetry breaking through the combined symmetry operation $[C_2 \mid C_{4_z}]$ (Figure 5d). This work represents a major conceptual advance, as it demonstrates that altermagnetism in frameworks can be designed a priori by mapping abstract symmetry criteria onto chemically accessible lattice architectures. Rather than relying on fortuitous material discovery, this methodology highlights chemistry as an active engine for symmetry realization, enabling the construction of molecular frameworks in which the required magnetic symmetry is encoded through ligand geometry and coordination motifs. Building on this symmetry-first design philosophy, recent work has demonstrated that AM symmetry in MOFs can not only be realized a priori, but also actively manipulated through coupling additional ferroic order parameters. In a particularly compelling example, Gu *et al.* introduced the concept of ferroelectric switchable altermagnetism, in which the AM spin splitting is directly coupled to the reversal of ferroelectric polarization.[67] Using spin-space group symmetry analysis combined with large-scale database screening, the authors identified a small set of candidate materials satisfying the stringent symmetry requirements for this coupling, including a hybrid improper ferroelectric Cr-based MOF as a representative case (Figure 5b). In this system, the AM state remains fully compensated in real space and nonrelativistic in origin, yet it becomes intrinsically coupled to ferroelectric polarization. As a result, the AM spin texture can be reversibly switched by electric-field-driven polarization reversal without modifying the underlying magnetic order parameter. This coupling between altermagnetism and ferroelectricity highlights a qualitatively new form of ferroic control, in which symmetry-engineered magnetic states in coordination frameworks can be actively manipulated through lattice degrees of freedom rather than magnetic fields.[68]

Most theoretical proposals for AM MOFs have so far focused on single-layer or strictly 2D coordination networks, where symmetry relations are local to the lattice plane. Extending altermagnetism beyond this limit, Che *et al.* introduced bilayer MOFs altermagnets as a qualitatively distinct platform in which interlayer symmetry becomes an active design degree of freedom.[69] In these systems, combined spatial and spin operations such as $[C_2 | R]$ and $[M_z | R]$ connect opposite spin sublattices that are spatially separated into different layers, while pure spatial inversion and mirror symmetries that would enforce spin degeneracy are selectively removed by the bilayer stacking sequence (Figure 5e). Because the surviving spin point group operations explicitly involve layer-resolved transformations, the AM electronic structure becomes sensitive to perturbations acting asymmetrically on the two layers. As a result, external electric fields can directly modulate the AM band structure through layer polarization, establishing bilayer coordination frameworks as a natural platform for electrically tunable altermagnetism.

**Building MOF altermagnets: chemical design strategies**

Building on the design strategies outlined above, altermagnetism in coordination frameworks can be rationalized in terms of a small set of chemically controllable figures of merit. Rather than being tied to specific compounds, these parameters capture the ability of MOFs to realize the symmetry operations and spin space groups required for AM order. All current theoretical proposals of AM frameworks can be understood within this compact design space. In this section, we identify four key parameters that govern the emergence and control of altermagnetism in framework materials: ligand symmetry, lattice architecture and dimensionality, magnetic sublattice differentiation and orbital design. Together, they define a chemically accessible map for designing AM coordination networks.

- Ligand symmetry

  Ligand symmetry provides the primary molecular origin of lattice symmetry in coordination frameworks. The local point symmetry, directionality and connectivity of an organic linker determine how coordination motifs propagate through space, thereby dictating the global symmetry of the extended network. As a result, small variations in ligand geometry or substitution pattern can modify rotational elements, mirror operations, or glide symmetries of the lattice without altering its overall topology. For example, the replacement of centrosymmetric linkers such as pyz by lower-symmetry heterocyclic ligands can reduce the symmetry of the metal-organic lattice and enable AM spin splitting.[63] This direct translation of molecular symmetry into lattice symmetry is a distinctive feature of coordination chemistry, allowing chemically similar frameworks to exhibit distinct spin space groups and AM anisotropies solely as a consequence of linker symmetry.

- Lattice architecture and dimensionality

    While ligand symmetry defines the local building blocks, the global network architecture determines which symmetry classes are accessible at the lattice level. By deliberately designing the underlying topology, coordination frameworks can be constructed to satisfy the symmetry requirements associated with AM order. For instance, square, rectangular, honeycomb, or Cairo-type lattices each support distinct rotational and symmetry elements that can stabilize different AM anisotropies.[50,51] Within this architectural freedom, dimensionality emerges as a powerful additional degree of control. 2D coordination networks naturally support planar rotational and glide symmetries that favor momentum dependent spin splitting, while multilayer and stacked architectures introduce interlayer symmetry operations that enable additional control channels, such as layer resolved spin textures. Fully 3D frameworks further expand the accessible design space by allowing nonsymmorphic symmetries and complex spin space group operations that are inaccessible in reduced dimensionality. In this context, changes in dimensionality are not merely geometric modifications, but constitute an active symmetry engineering strategy that qualitatively alters the nature and tunability of AM states.

- Magnetic sublattice differentiation

    A central structural requirement for altermagnetism is the presence of magnetically compensated yet crystallographically inequivalent spin sublattices. Unlike conventional antiferromagnets, where opposite spins occupy symmetry equivalent sites related by simple spatial operations, AM frameworks require antiparallel spins to reside on distinct crystallographic positions with different local environments. This controlled inequivalence prevents the enforcement of spin degeneracy while preserving global magnetic compensation. Coordination frameworks provide several chemical routes to achieve this condition. Structures containing multiple metal sites per unit cell, mixed coordination geometries, or the coexistence of metal centered and ligand centered spin carriers can place spins on distinct Wyckoff positions or on sites of reduced symmetry.[27] Mixed valence frameworks, heterometallic lattices and networks with inequivalent coordination nodes offer particularly natural pathways to this type of spin differentiation.

- *Orbital design and energy scales*

    Beyond structural symmetry, orbital design governs the electronic energy scales that determine whether AM order becomes experimentally observable. Controlling the symmetry, spatial distribution and energetic alignment of ligand frontier orbitals, frontier molecular orbital engineering is a key tool to enable the deliberate reorganization of exchange pathways and spin delocalization across the lattice. This orbital control can induce qualitative changes in magnetic order, including the stabilization of AM ground states and the switching between distinct AM anisotropies within the same structural platform.[65] In addition to selecting symmetry allowed states, orbital design sets the magnitude of exchange interactions and spin splitting. Ligand centered spin polarization, nonbonding molecular orbitals and redox active

frameworks enhance magnetic coupling and promote extended spin delocalization, leading to higher magnetic ordering temperatures and larger electronic spin splittings. In this way, orbital engineering complements symmetry design by transforming altermagnetism from a symmetry allowed possibility into an energetically robust and potentially functional magnetic phase.

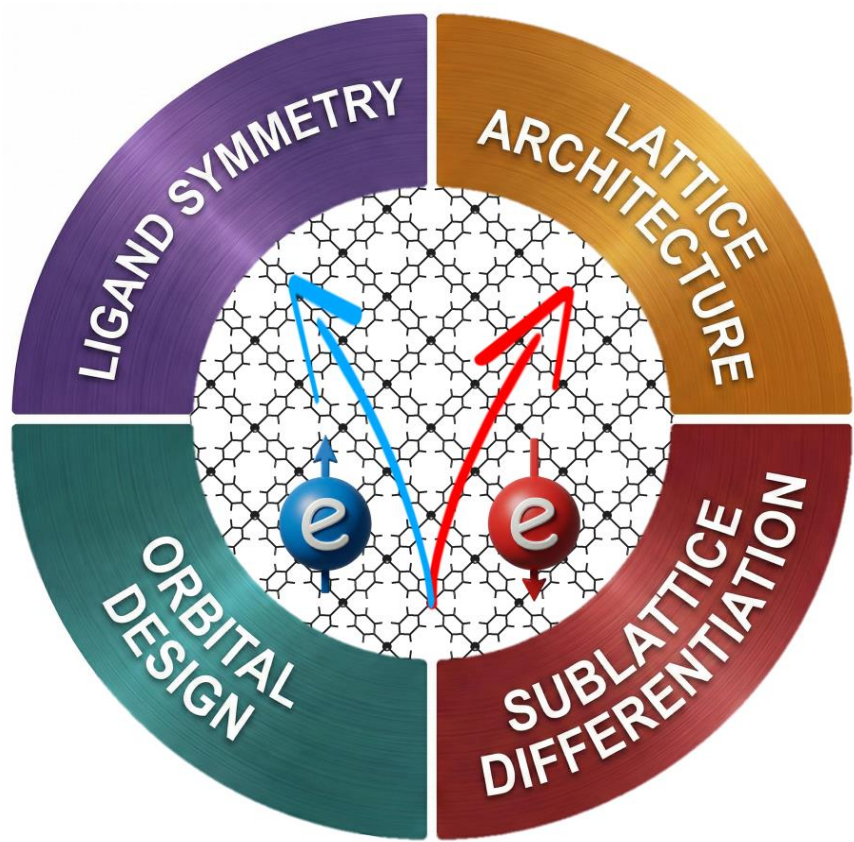

**Figure 6:** Schematic representation of discussed strategies for inducing altermagnetism in MOFs.

Taken together, these four figures of merit establish altermagnetism in coordination frameworks as a designable and programmable property rather than a material-specific curiosity (Figure 6). The modular nature of framework chemistry enables the construction of entire families of AM materials in which symmetry, sublattice structure and orbital character can be independently tuned. This Perspective shifts the search for altermagnets from empirical discovery toward rational design, opening a pathway to chemically engineered AM platforms with controllable functionality and relevance for future spintronic and quantum technologies.

# ■ CHALLENGES, PROSPECTS AND FUTURE DIRECTIONS

## Fundamental challenges

Despite the rapid conceptual and theoretical advances outlined above, the realization of altermagnetism in MOFs poses a distinct set of challenges that extend beyond conventional magnetic materials design. These challenges arise from the requirements that altermagnetism imposes on structural fidelity, experimental observables and energetic robustness. In the following, we critically assess the most relevant obstacles that must be addressed to translate symmetry-engineered AM concepts into experimentally accessible framework materials, while highlighting emerging experimental capabilities that make this goal increasingly realistic.

A central requirement underlying many of these challenges is the ability to access framework materials with well-defined structure and sufficient sample quality for advanced measurements. Encouragingly, several experimental advances directly address the challenges associated with identifying altermagnetism in coordination frameworks. The realization of electrically conductive MOFs as single crystals enables access to intrinsic, symmetry-resolved electronic and transport properties without averaging effects from grain boundaries, which is essential for detecting anisotropic and momentum-dependent responses.[70,71] In parallel, the successful integration of MOFs into field-effect transistor (FET) architectures demonstrates their compatibility with device-level measurements, providing electrostatic control over carrier density and enabling gate-tunable transport signatures relevant to AM phases.[21,72] Bottom-up fabrication strategies, including layer-by-layer and surface-assisted growth, further allow the preparation of oriented thin films with controlled thickness and crystallographic alignment, opening realistic routes toward angle-dependent transport and spectroscopic probes.[73] Finally, the use of chemically programmable molecular building blocks permits the deliberate encoding of symmetry, orbital character and magnetic functionality at the molecular level, facilitating the rational construction of framework materials tailored for advanced magnetic and electronic characterization. Together, these developments indicate that the experimental tools required to probe altermagnetism in MOFs are rapidly converging toward practical implementation.

A further fundamental challenge lies in the experimental identification of altermagnetism itself. Because AM phases are fully compensated in real space, conventional magnetometry is inherently insufficient to distinguish them from collinear AFM order. Instead, altermagnetism manifests through symmetry-protected, momentum-dependent spin splitting and associated anisotropic electronic responses.[2] In other material classes, these signatures have been accessed using momentum-resolved spectroscopies, symmetry-sensitive transport measurements and nonlinear electronic responses, providing clear experimental benchmarks beyond net magnetization.[74–77] Importantly, several of these experimental approaches have already been successfully implemented in metal-organic frameworks and related coordination networks. Angle-resolved photoemission spectroscopy (ARPES) has been applied to surface-supported 2D MOFs and coordination networks to directly probe their electronic band structure and metal-ligand hybridization, demonstrating that momentum-resolved measurements are feasible in these systems when prepared with sufficient structural order.[78,79] In parallel, electrically conductive and magnetic MOFs have been integrated into transport measurements, including anisotropic conductivity and Hall-effect experiments, establishing a precedent for probing symmetry-dependent electronic responses in framework materials.[70] Indeed, these results indicate that the experimental toolbox required to identify altermagnetism is not fundamentally incompatible with coordination frameworks but rather requires the extension of already demonstrated methodologies to symmetry-engineered AM MOFs.

A final fundamental challenge concerns the thermal stability and energetic robustness of AM phases. As summarized in Table 1, theoretical proposals of AM MOFs predict momentum-dependent spin splittings typically in the range of 10 – 300 meV. While these values are already substantial and comparable to many two- and three-dimensional inorganic altermagnets, they remain significantly below the largest reported nonrelativistic AM spin splittings, which reach values close to ~1 eV in inorganic systems such as CrSb.[77] This gap highlights an important challenge for framework-based altermagnets, namely the need to further enhance electronic energy scales through stronger exchange interactions and increased spin delocalization. In parallel, a marked disparity persists in magnetic ordering temperatures. Whereas altermagnetism in inorganic materials has been identified in systems exhibiting magnetic order up to and including room temperature,[76] predicted ordering temperatures for MOF-based altermagnets generally do not exceed ~200 K. More broadly, within magnetic MOFs, transition temperatures approaching ambient conditions have only been achieved in frameworks incorporating redox-active ligands, where metal-radical exchange dominates,[43] while systems relying primarily on metal-metal superexchange typically order at cryogenic temperatures. These considerations underscore that achieving AM MOFs combining elevated magnetic ordering temperatures with spin splitting approaching the energy scales of inorganic altermagnets represents a central challenge for the maturation and practical relevance of the field.

Taken together, these challenges highlight that the realization of altermagnetism in coordination frameworks is not limited by conceptual feasibility, but by the integration of structural precision, symmetry fidelity, experimental accessibility and energetic robustness within a single material platform. Importantly, each of these challenges is already being actively addressed within the broader MOF community, suggesting that the remaining barriers are technical rather than fundamental. As a result, the emergence of experimentally verified AM frameworks should be viewed not as a distant prospect, but as a realistic next step enabled by continued advances in reticular chemistry, materials synthesis and characterization.

**New horizons: emerging directions and unexplored territories**

Among the emerging directions for altermagnetism in coordination frameworks, layer stacking represents one of the most immediately impactful and chemically accessible strategies. Notably, many of the magnetic MOFs discussed throughout this Perspective adopt intrinsically layered, vdW-bonded structures, where individual magnetic layers are weakly coupled along the stacking direction. This structural motif naturally enables stacking engineering as an additional degree of freedom. By controlling stacking sequence or symmetry relations between adjacent layers, it becomes possible to generate or modify the spin space group operations required for altermagnetism without altering the in-plane lattice. Exploiting the layered nature of magnetic MOFs in this way provides a powerful and largely unexplored route to realize AM phases through interlayer symmetry design rather than purely 2D lattice engineering.

Intercalation provides a closely related and experimentally validated route to manipulate altermagnetism in coordination frameworks.[80] Importantly, several experimental studies have already demonstrated that the intercalation of metallic atoms or molecular species into layered MOFs can induce dramatic changes in both magnetic and electronic properties, including enhanced magnetic ordering temperatures, modified exchange interactions and substantial increases in electrical conductivity.[43,44] These results establish intercalation as a powerful post-synthetic strategy to tune framework properties without altering the underlying lattice topology. Notably, in layered vdW inorganic materials, atomic intercalation has recently been shown to induce or stabilize AM phases by modifying interlayer symmetry relations and electronic filling. Intercalation of interfacial Co atoms in the layered structure of $NbSe_2$ has shown to introduce AM spin splitting confirmed experimentally via spin-polarized ARPES in hybrid $CoNb_4Se_8$.[81] This precedent strongly suggests that analogous strategies in layered MOFs could be exploited to engineer altermagnetism through controlled interlayer coupling and symmetry manipulation. Building on these experimental foundations, intercalation emerges not as a speculative concept, but as a realistic and versatile tool for tailoring symmetry-driven magnetic behavior in coordination frameworks.

Beyond chemical design, external stimuli offer a powerful and largely untapped route to manipulate magnetic order in MOFs. In particular, hydrostatic pressure has already been demonstrated to dramatically tune magnetic interactions in layered magnetic materials by modifying interlayer spacing, orbital overlap and charge redistribution without altering the underlying lattice symmetry.[82] Recent experiments show that pressure can strongly enhance interlayer exchange, coercivity and magnetic robustness while preserving crystallographic symmetry and long-range order, establishing pressure as an effective control knob for magnetic energy scales in MOFs.[83–85] Remarkably, $Li_{0.7}Cr(pyz)_2Cl_{0.7}$ layered magnetic MOF has shown pressure tuning with a coercivity coefficient up to 4 kOe/GPa, which represents a powerful platform for manipulating the magnetic properties of molecular frameworks.[86] Closely related to pressure, mechanical strain provides an additional and highly promising handle. In inorganic materials, strain has already been shown to induce and control AM phases through symmetry lowering and band structure reconstruction.[87] Interestingly, $ReO_2$ conventional antiferromagnet has been predicted to possess a strain-induced AFM-to-AM transition, which is dominated by a prominent d-wave AM anisotropy.[88] In coordination frameworks, strain is known to strongly affect spin state energetics and has been used to trigger spin crossover and spin transition phenomena,[89] highlighting the pronounced sensitivity of MOF electronic structure to lattice deformation. By extension, pressure and strain driven modulation of interlayer coupling and orbital hybridization represent highly promising strategies to access or stabilize AM phases, particularly in vdW stacked frameworks where symmetry protected spin splitting is sensitive to subtle structural changes. Importantly, these approaches parallel well-established stimulus controlled magnetic phase engineering in inorganic layered magnets, suggesting clear and experimentally realistic pathways to explore altermagnetism beyond purely chemical tuning.

An additional and largely unexplored direction concerns the realization of metallic AM MOFs. To date, all reported AM framework proposals correspond to semiconducting systems, in which the Fermi level lies within the band gap. This electronic structure requires external tuning, such as electrostatic gating or chemical doping, to shift the Fermi level toward the valence or conduction bands in order to experimentally access spin dependent transport or momentum resolved signatures.[21] In contrast, intrinsically metallic AM frameworks would host symmetry driven spin splitting directly at the Fermi surface, enabling straightforward electrical detection and device integration. The development of metallic AM MOFs therefore represents a major opportunity for the field, combining the symmetry design principles of coordination chemistry with the transport functionality required for practical spintronic applications.

Heterostructure engineering represents another highly promising route for realizing and controlling altermagnetism in low-dimensional magnetic materials.[90] In inorganic layered materials, the creation of vdW heterostructures has already been shown to induce AM states through interfacial symmetry breaking, band rearrangement and proximity effects, enabling electric-field control and ferroic switching of AM spin splitting.[91–93] These advances provide a compelling blueprint for MOF-based systems, where substrate coupling has been predicted to drive magnetic phase transition due to charge transfer and strain.[94] Importantly, experimental heterostructures combining 2D MOFs with inorganic 2D materials have already been demonstrated, establishing the feasibility of integrating molecular magnets into layered hybrid architectures. For example, a monolayer of magnetic Cu-dicyanoanthracene has been deposited in the surface of the superconductor $NbSe_2$, which has shown to exhibit different structural arrangements which could be further coupled with magnetic and superconducting properties.[95] In such systems, interfacial charge transfer, electrostatic coupling and symmetry lowering at the heterointerface offer powerful handles to activate or modulate altermagnetism without altering the intrinsic framework chemistry. This positions MOF-based heterostructures as a natural extension of current AM design strategies, combining molecular-level tunability with interfacial band and symmetry engineering.

Twist engineering has become a well-established strategy in inorganic vdW materials, where relative rotational misalignment between layers provides a powerful handle to modify symmetry, electronic structure and collective quantum states.[96] In contrast, this approach has remained essentially unexplored in MOFs, despite the fact that many layered MOFs are held together by weak vdW interactions and are therefore, in principle, amenable to similar manipulations. Recently, the first experimental demonstration of twist engineering in 2D MOFs has been reported, showing that mechanically exfoliated

MOF layers can be deterministically stacked with controlled relative orientation, leading to pronounced modifications of anisotropic physical properties.[97] Although demonstrated so far in the context of optical anisotropy, this work establishes the experimental feasibility of twist-controlled MOF heterostructures. Extending this concept to magnetic and electronically active frameworks opens a largely unexplored avenue for inducing or tuning AM symmetries through rotational control, directly leveraging the sensitivity of altermagnetism to lattice and interlayer symmetry relations.

Finally, coordination frameworks offer a unique opportunity to explore altermagnetism in quasi-1D architectures. Recent theoretical work has demonstrated that AM states can arise in assemblies of parallel magnetic chains, where the relative arrangement, interchain coupling and symmetry relations between chains play a decisive role.[98] Crucially, many MOFs are already known to host highly anisotropic magnetic lattices, featuring well-defined quasi-1D chains that are weakly coupled through organic linkers or vdW interactions.[97,99] These materials naturally realize the key ingredients identified in 1D AM proposals, namely spatially separated chains with tunable interchain spacing and magnetic coupling. As such, MOFs provide a chemically accessible and structurally versatile platform to embed 1D altermagnetism within an extended crystalline host, offering a promising bridge between recent theoretical predictions and experimentally realizable low-dimensional AM systems.

## ■ CONCLUSIONS AND OUTLOOK

The emergence of altermagnetism opens a new frontier in magnetic materials design, where symmetry engineering becomes a central guiding principle. In this Perspective, we discuss that MOFs provide a uniquely powerful and timely platform to realize, generalize and ultimately control altermagnetism through chemical design. Unlike conventional inorganic magnets, coordination frameworks allow symmetry, lattice topology and electronic structure to be encoded at the molecular level, transforming symmetry from a fixed constraint into an actively tunable design parameter.

By surveying recent theoretical proposals and experimental advances, we have shown that the symmetry requirements for altermagnetism can be fulfilled across a broad range of framework architectures, encompassing 3D bulk frameworks, layered vdW-bonded materials, genuinely 2D networks and emergent quasi-1D motifs. More importantly, we have emphasized that the modular nature of MOFs enables systematic exploration of AM design principles beyond isolated model lattices, opening access to stacking engineering, intercalation, external stimuli, heterostructures and orbital-level control strategies that are largely inaccessible in dense inorganic solids. These opportunities position coordination frameworks as a platform to expand its phenomenology and functional scope.

The challenges described in this Perspective are not fundamental barriers, but technical and materials-oriented questions that are already being addressed within the rapidly evolving MOF community. The convergence of advances in reticular chemistry, electronic structure engineering and symmetry-sensitive characterization techniques suggests that the experimental realization and control of AM states in coordination frameworks is feasible. We therefore anticipate that magnetic MOFs will play a central role in shaping the next phase of altermagnetism research, bridging symmetry-driven magnetic order with chemically programmable quantum materials and device-oriented functionalities.

## AUTHOR INFORMATION

Corresponding author

*E-mail: j.jaime.baldovi@uv.es

Author contributions

The manuscript was written through contributions of all authors. All authors have given approval to the final version of the manuscript.

Notes

The authors declare no competing financial interest.

## ACKNOWLEDGEMENTS

The authors acknowledge financial support from the European Union (ERC-2021-StG101042680 2D-SMARTiES), the Spanish MCIU (PID2024-162182NA-I00 2D-MAGIC), Spanish MICINN (Excellence Unit "Maria de Maeztu" CEX2024-001467-M) and the Generalitat Valenciana (grant CIDEXG/2023/1).